\documentclass{raa}

\usepackage{array}
\usepackage{booktabs}
\usepackage{graphicx, times}
\usepackage{natbib}
\usepackage{amssymb,amsmath}
\usepackage{bm}
\usepackage{xcolor}

\begin{document}

\title{Stellar Dynamical Modeling - Counting Conserved Quantities}


\author{Richard J. Long\inst{1,2}, Shude Mao\inst{3,4}, Yougang Wang\inst{4,5}}


\institute{Department of Astronomy, Tsinghua University, Beijing 100084, China; {\it rjlastro@yahoo.com; Orcid:0000-0002-8559-0067} \\
        \and
           Jodrell Bank Centre for Astrophysics, Department of Physics and Astronomy, The University of Manchester, Oxford Road, Manchester M13 9PL, UK\\
        \and
        	Department of Astronomy, Tsinghua University, Beijing 100084, China\\
        \and
        	National Astronomical Observatories, Chinese Academy of Sciences, Beijing 100101, China\\
        \and
        	School of Astronomy and Space Science, University of Chinese Academy of Sciences, Beijing 100049, China\\
           }

\date{Received~~2022 month day; accepted~~2022~~month day}

\renewcommand{\labelitemi}{$\bullet$}
\newcolumntype{M}[1]{>{\centering\arraybackslash}m{#1}}

\abstract{
Knowing the conserved quantities that a galaxy's stellar orbits conform to is important in helping us understand the stellar distribution and structures within the galaxy.  Isolating integrals of motion and resonances are particularly important, non-isolating integrals less so.  We compare the behavior and results of two methods for counting the number of conserved quantities, one based on the correlation integral approach and the other a more recent method using machine learning.  Both methods use stellar orbit trajectories in phase space as their only input, and we create such trajectories from theoretical spherical, axisymmetric and triaxial model galaxies.  The orbits have known isolating integrals and resonances.  We find that neither method is fully effective in recovering the numbers of these quantities, nor in determining the number of non-isolating integrals.  From a computer performance perspective, we find the correlation integral approach to be the faster.  Determining the algebraic formulae of (multiple) conserved quantities from the trajectories has not been possible due to the lack of an appropriate symbolic regression capability.  Notwithstanding the shortcomings we have noted, it may be that the methods are usable as part of a trajectory analysis tool kit.
}

\authorrunning{Richard J. Long et al.}            
\titlerunning{Counting Conserved Quantities}  

\maketitle


\section{Introduction}
\label{sec:intro}
In this Introduction, we first consider briefly traditional stellar dynamical modeling, and then point out how some aspects of the modeling may be dealt with using machine learning techniques.  We complete the Introduction by setting our objectives for the investigation described in this article.  We describe our approach to meeting the objectives in the next section.

The stellar dynamics of galaxies are explored by developing some model of a galaxy, constraining it with observational data, and then examining the model to see what might be learnt about the real galaxy.  \cite{BT2008} describe the theory behind many of the modeling techniques that might be used. 
A key point which must be understood about external galaxies is that our models are only indicative and illustrative of real (external) galaxies because of the current technical limitations of the instruments we employ for observations: in essence, we can not obtain 3D galaxy data.  Observing and modeling our Galaxy is different as we do have the capability to collect 3D spatial and velocity data.

Theoretically, the collisionless Boltzmann equation (CBE), by linking a galaxy's gravitational potential and its phase space distribution function $f(\textbf{x},\textbf{v},t)$, gives us a start point for modeling. Integrating the distribution function (a probability density function) in various ways gives us a means of creating model observables which can then be compared with real stellar observables (that is, the observed data). Equations of motion (if they are needed) come from Hamiltonian mechanics, giving a 6D phase space.  Constants or integrals of motion come from considering the Jeans theorems, and lead to isolating integrals which partition phase space, and non-isolating integrals which do not. Typical isolating integrals are energy, and some or all of the components of angular momentum. Spatial resonances may arise from the equations of motion under certain circumstances, and resonant orbits may have trajectories which influence the shape sub-structures present in the galaxy.
Steady state where the potential and distribution function are not time dependent is a major simplifying assumption which is often used.  In a similar vein, the form of the potential is frequently assumed from the shape of the galaxy.

Modeling methods typically come from the Jeans equations, or from processes trying to avoid knowing the distribution function, or from using processes where the form of the distribution function is assumed.  Schwarzschild's method \citep{Schwarz1979} and the made-to-measure (M2M) method of \cite{Syer1996} are two methods which use weighted orbits or particles to tailor models with the intention that, by adjusting the weights, model observations will match real observations.  While the weights can be interpreted astrophysically their values are in fact set by purely numerical methods \citep{Long2016, Long2018, Long2021}.  The M2M method we will refer to later.  Jeans equation methods may yield non-physical models \citep[for example,][]{vandeVen2003, Cappellari2008} so we will not discuss them further.

Having set the context using a more traditional stellar dynamical modeling approach, we now consider what has been achieved and might be achieved by using machine learning.  Bulk processing of noisy, image data at radio and optical wavelengths to classify galaxies is well-advanced - see, for example, \cite{Clarke2020}, \cite{Canducci2022} and \cite{Tang2022}.  For single galaxies, orbit trajectory data and normalizing flows provide the capability to determine distribution functions and accelerations from the gravitational potential - see \cite{Green2021}, \cite{An2021} and \cite{Naik2022}.  \cite{An2021} used a \cite{Hernquist1990} model in their experiments. From our own experiences (as yet unpublished) converting to other models appears to work satisfactorily.  Training models to learn and follow conservation laws in a Hamiltonian context was accomplished by \cite{Greydanus2019} with a Lagrangian implementation in \cite{Cranmer2020}.  Data driven discovery of coordinates and equations of motion using an autoencoder architecture with symbolic regression was investigated by \cite{Champion2019}.  Symbolic regression \citep[see review by][]{Cava2021} is concerned with taking numerical input and output data and producing an algebraic formula to convert the input to the output.  Its value in a stellar dynamical modeling context is appreciable: it means, for example, that we should be able to take the weights from a M2M modeling run and produce an algebraic formula for the galaxy distribution function.  Both \cite{Nature2021} and \cite{Meng2022} deal with physics-informed machine learning with the latter surveying the methods and techniques available.  What is required in due course is that the 
physics-informed techniques are investigated for their applicability in an astronomy, stellar dynamics context.

What is hopefully now clear is that there is sufficient overlap between the traditional stellar dynamical modeling techniques and machine learning techniques to warrant further investigation.
The end position we should be aiming for is to see how far we can get in developing new stellar dynamical modeling tools based on machine learning to augment if not replace the existing traditional tooling.  The gap is not that large in that appropriate machine learning based tools are emerging, and the existing traditional M2M scheme already has some features found in neural networks, for example back propagation \citep{Rumelhart1986}.

Turning now to this paper, as we said in the Abstract, knowing the conserved quantities that a galaxy's stellar orbits conform to is important in helping us understand the stellar distribution and structures within the galaxy.  Isolating integrals of motion and resonances are particularly important, non-isolating integrals less so.  The first step is knowing just how many conserved quantities an orbit conforms to, and we will do this by examining and comparing two methods for counting such quantities, the first based on a traditional approach and the second on machine learning.  The traditional method is closely related to the correlation integral method described in \cite{Carpintero2008} which is based on the work of \cite{Carnevali1984} and \cite{Barnes2001}, which in turn are based on \cite{Grassberger1983}.  The machine learning method is as described in \cite{Liu2021}, and takes a manifold dimensionality approach enabled by \cite{Saremi2019}.  \cite{Liu2021} was developed in a physics rather than an astronomy context, and we are not aware of any previous galaxy-based usage.
As a consequence, our research objectives are to compare and contrast methods (with and without the exploitation of machine learning) for determining the number of integrals of motion and resonances from galaxy stellar orbit trajectories.  These trajectories will be created by integrating the equations of motion in various theoretical gravitational potentials representing galaxies, where the expected number of isolating integrals is known in advance.

The structure of our paper is as follows.  Section \ref{sec:approach} describes at a top level the approach we will take to our investigations.   Relevant theory and descriptions of the methods are in Section \ref{sec:theory}. Our results and subsequent discussion are in Sections \ref{sec:results} and \ref{sec:discussion}, with our conclusions in Section \ref{sec:conclusions}.

\section{Approach} 
\label{sec:approach}

The motion of stellar objects within a galaxy is usually described and analyzed mathematically by using Hamiltonian mechanics \citep{BT2008}.
Galaxy stellar orbit phase space trajectories define manifolds \citep{Arnold1989}, and we use the local properties of manifolds to help us determine 
the number of conserved quantities present in an orbit trajectory.  In particular, manifolds are locally homeomorphic to a Euclidean space, and conserved integral quantities act to reduce the effective dimensionality of the manifold, as do resonances.  As indicated in the Introduction, Section \ref{sec:intro}, we are comparing two
different methods, a recent method
by \cite{Liu2021} using manifold  machine learning techniques, and an earlier, correlation integral method based on the work of \cite{Grassberger1983}.  
\cite{Liu2021} in their supplementary material do mention briefly the correlation integral method but refer to it as the `fractal' method.  In this 
paper, our correlation integral methodology is mainly influenced by \cite{Carpintero2008}.

Our comparison utilizes theoretical galaxy models of increasing complexity, where the normally expected number of isolating integrals is known.  If more than the expected number are found,  this implies that a trajectory has additional conserved quantities, for example, resonances \citep{Carpintero2008}.  If fewer are found, then we are dealing with irregular or chaotic orbits.  In 6D phase space, the minimum number of conserved quantities is one (corresponding to energy) and the maximum is five (which would imply a 1D orbit).  It is important to note that the methods operate with the trajectory of a single orbit.  Even though we will talk about galaxy models and orbits from those models, in no sense do the methods operate at a galaxy level utilizing many orbits in one model: the methods operate on single orbits only.

Our galaxy models have various morphologies and include spherical, axisymmetric (oblate), and triaxial (ellipsoidal) theoretical models.  We have deliberately picked models that have been used elsewhere for similar related work.  Our spherical models are taken from \cite{Plummer1911} and \cite{Hernquist1990}, and our axisymmetric model is the logarithmic model from \cite{Richstone1980} in both cored and singular forms.  For our triaxial model, we use the perfect ellipsoid described in \cite{deZeeuw1985}. 

Both methods take orbit trajectories as their only data input.  We create 6D phase space orbit trajectories in Cartesian coordinates, but also use coordinates specific to the morphologies of our galaxy models to help us shed light on the functioning of the methods..  For spherical models, we use 2D Cartesian configuration coordinates in the individual orbital planes of the trajectories, and similarly for axisymmetric models where we use the meridional plane.  For triaxial models, we also use 3D confocal ellipsoidal coordinates.  For operational, time efficiency reasons, we limit our work to 64 orbit trajectories per galaxy model.  The trajectories are the same for both modeling schemes.  Note that resonances, if present, may manifest themselves differently in different coordinate systems \citep{Papa1996}.

For each model within the three galaxy morphologies mentioned, we execute four modeling runs made up of the two methods with two coordinate schemes, all utilizing the same 64 orbits.  In the comparison, we are looking for a high level of consistency between the two methods with the expectation that the number of conserved quantities for an individual orbit will be correct and the same (within, of course, the limitations of the analysis techniques available to us).
  
Our work is limited to a comparison between methods using trajectory data from simple theoretical models. We are concerned to understand whether the individual methods function correctly and consistently with 6D phase space data.  As a consequence, more complex scenarios, such as the following, are out of scope,
\begin{itemize}
\item perturbed trajectories,
\item general time-varying systems where the gravitational potential is a function of time,
\item time-averaged conserved quantities as in \cite{Qin2021} for example,
\item rotating bar potentials as in \cite{Barnes2001}, and
\item the impact of adiabatic variations on orbits.  
\end{itemize}
In addition, determining formulae for conserved quantities is not performed: various schemes are described in the literature but, from our own experiences with them, no robust, reliable method exists as yet to do so for multiple conserved quantities in 6D phase space.

For the remainder of this paper, we refer to the machine learning method as the ML method, and the more traditional method using the correlation integral as the CI method.  To clarify, since we are not able to determine formulae and therefore are unable to confirm quite what we have counted, we use the term `conserved quantities' (CQ) to cover both integrals and resonances.  This lack of formulae is not specific to our work, and is present in earlier CI work.  We use the notation $CQ = n$ to indicate $n$ conserved quantities.

To be clear, our work is concerned with stellar dynamics in a galaxy context.  Use of methods and the results achieved must be interpreted in that context: they may not be applicable in other contexts.

\section{Theory and Methods} 
\label{sec:theory}

\subsection{The CI Method}
\label{sec:CImethod}

Our design and implementation of the correlation integral (CI) method is based on that in \cite{Carpintero2008}.  
The method aims to determine the number of effective dimensions $N_{\rm eff}$ of an orbit.
Having done that the number of conserved quantities $CQ$ for the orbit is just 
\begin{equation}
\label{eqn:CQeqn}
    CQ = N_{\rm PS} - N_{\rm eff},
\end{equation}
where $N_{\rm PS}$ is the number of phase space dimensions of the orbit.

Thinking now of the manifold the phase space orbit trajectory defines, given the number of conserved quantities reduces the
dimensionality of the trajectory, if we define a hypersphere of radius $r_{\rm s} << 1$ around a trajectory point,
the number of other trajectory points $N_{\rm pts}$ in the hypersphere should increase as $r_{\rm s} ^{N_{\rm eff}}$ as $r_{\rm s}$ is increased.
So, for increasing $r_{\rm s}$
\begin{equation}
\label{eqn:prop}
    N_{\rm pts} \propto r_{\rm s} ^{N_{\rm eff}},
\end{equation}
a power law relationship which allows $N_{\rm eff}$ and thus $CQ$ to be determined.  

More formally, the correlation integral can be written as
\begin{equation}
\label{eqn:formal}
	C(r_{\rm s}) = \lim_{N \to \infty} \frac{2}{N(N-1)} \sum ^N_{i=1} \sum^N_{j>i} \Theta(r_{\rm s} - |\mathbf{X}_i - \mathbf{X}_j|),
\end{equation}
where $\mathbf{X}_i$ and $\mathbf{X}_j$ are phase space trajectory points, $\Theta$ is the Heaviside function, and $N$ is the number of trajectory points contributing to the integral.   Changing terminology slightly, it should be clear that the correlation integral is just the 2-point correlation function of the trajectory points.  Taking Equations (\ref{eqn:prop}) and (\ref{eqn:formal}) together, $N_{\rm pts}$ is equivalent to $C(r_{\rm s})$.  From Equation (\ref{eqn:formal}), the single computer processor performance of the method nominally scales as $N^2$ but the $C(r_{\rm s})$ calculation is readily amenable to parallelization: in computing terms, it is just a \textit{for} loop within a \textit{for} loop.

For $r_{\rm s}$, from experimentation, we take $80$ values uniformly spaced logarithmically
in the range [$10^{-3}$, $1$].  In subsequent sections, we will refer to $r_{\rm s}$ as a scale length.
Prior to calculating the correlation integral, we normalize the coordinates of a trajectory's points using min-max normalization as in \cite{Floss2018}.
Distances $d$ between two points in phase space (on the manifold) are calculated as 
\begin{equation}
	d = \sqrt{\delta x^2 + \delta y^2 + \delta z^2 + \delta v_x^2 + \delta v_y^2 + \delta v_z^2},
\end{equation}
where the difference in positions is $(\delta x,\delta y,\delta z)$ and in velocities $(\delta v_x,\delta v_y,\delta v_z)$.

We utilize two approaches in determining an integer value for $N_{\rm eff}$ and subsequently $CQ$.  We can work with the correlation integral directly, or we can work with its (numerically determined) gradient.  We have implemented and use both approaches but our preference is the gradient approach.  All the CI results described here are based on that approach, with the direct approach being used in a supporting role. 

$N_{\rm eff}$ is the gradient in a log-log formulation of Equation (\ref{eqn:prop}). This gradient varies with $r_{\rm s}$, is not a constant, and does
not have an integer value that we can immediately associate with $N_{\rm eff}$.  We arrive at a constant integer value by fitting possible
integer values ($1$ to $5$) of the gradient, and taking the value with the smallest squared residual with the integral gradient.  We do not use all the gradient
points in a single fitting process but using a number of successive gradient points (eight in our case), and perform multiple rolling fits
of the gradient by advancing the start point by one position along the gradient for each fit.  As part of the fitting process, we employ a range constraint on the selected points to deal with binning noise in constructing the correlation integral and its gradient: the minimum to maximum difference has to be less than some tolerance ($0.2$ initially).  If no gradient can be found using a given tolerance, the tolerance is relaxed slightly (by $0.05$) until integer gradient is found.

The approach using the correlation integral directly also employs the rolling fit tactic but fits a straight line to the integral points themselves, and uses the line's gradient to arrive at a value for $N_{\rm eff}$.  \cite{Carpintero2008} use a similar process.

\begin{figure}[h]
    \centering
    \caption{CI Method - example plots}
	\label{fig:ciplots}
    \begin{tabular}{cc}   	
      \includegraphics[width=55mm]{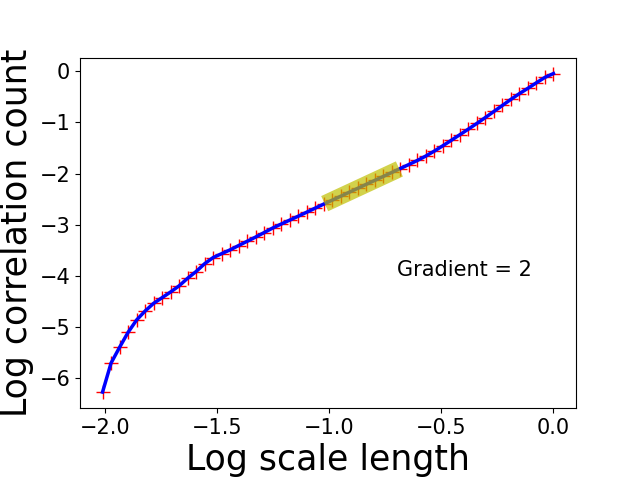} &  \includegraphics[width=55mm]{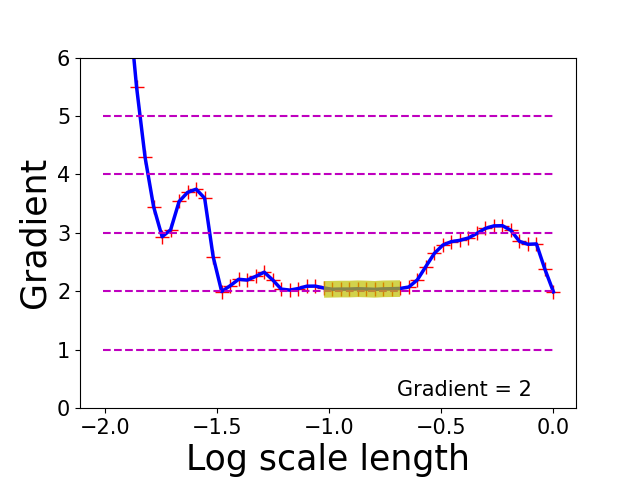} \\
    \end{tabular}
    
\medskip  
Example plots from using the CI method on trajectory data taken from a triaxial model orbit. The left hand plot shows the correlation integral plotted against scale length as a log-log plot.  The right hand plot shows the gradient of the curve on the left with the points contributing to the best-fitting integer gradient ($2$) marked in yellow.  In this case, the gradient of $2$ corresponds to the number of conserved quantities being $4$.  The fitting process is described in Section \ref{sec:CImethod}.
\end{figure}

Figure \ref{fig:ciplots} illustrates the CI method for an orbit and contains two plots, one showing the correlation integral, and a second showing the best-fitting integer gradient.  The points contributing to the best-fitting gradient are highlighted in yellow on both plots. 6D phase space is being used so, in this case, a gradient of $2$ corresponds to the number of conserved quantities being $4$.

\subsection{The ML method}
\label{sec:MLmethod}
Functionally, the ML method has similar aims to the CI method but whereas the CI method attempts to determine the number of effective dimensions of an orbit trajectory, the ML method attempts to determine the number of conserved quantities an orbit trajectory conforms to.  Using Equation (\ref{eqn:CQeqn}) from the previous section, the ML method obtains a value for $CQ$ and then $N_{\rm eff}$ .  Cross-matching mathematical symbols with \cite{Liu2021}, their $\hat{s}$ is the equivalent of our $N_{\rm eff}$, and their $n_{\rm eff}$ is our $CQ$.

Since we use their main algorithm and software largely unchanged, we only briefly outline the method and point the reader to \cite{Liu2021} and \cite{Saremi2019} for a fuller explanation.  The method has three main phases, 
\begin{enumerate}
\item preprocessing involving trajectory data whitening, and (optional) removal of any linear conserved quantities using Principal Component Analysis (PCA),
\item trajectory manifold sampling using a neural empirical Bayes technique \citep{Saremi2019} to characterize the local tangent space by perturbing the trajectory data, and
\item use of PCA explained variance ratios to estimate the number of effective dimensions and the number of conserved quantities.
\end{enumerate}
It is important to note that phase (2) above is repeated multiple times using different degrees of perturbation controlled by a scale length parameter.

We use the AI Poincar\a'e Python software package produced by \cite{Liu2021} unchanged except for making the user interface more amenable to batch operation, and splitting the package so that the plots are produced in a script separately runnable from the main modeling code. The original package is publicly available on the internet (github.com/KindXiaoming/aipoincare).   In training the neural networks, we use the `Adam' optimizer \citep{Adam2014} and mean square error for the loss function.
In machine literature generally, there is no precise guidance on setting the number of hidden layers and the number of nodes in neural networks other than that more complex modeling requires more hidden layers and nodes. After some experimentation, we find this to be true in our work where we use two layers of 128 nodes for spherical models, four layers of 192 nodes for our axisymmetric model, and four layers of 320 nodes for the triaxial model.  Similarly, we experiment with the values of the training learning rate and the number of training iterations, and arrive at a value of $0.001$ for the learning rate and values of 500 (spherical models) and 2000 (other models) for the training iterations.  We take values of $[0.001, 0.01, 0.025, 0.05, 0.075, 0.1, 0.15, 0.25, 0.5, 1.0]$ for the scale (perturbation) lengths.

\begin{figure}[h]
    \centering
    \caption{ML Method - example plots}
	\label{fig:mlplots}
    \begin{tabular}{ccc}   	
      \includegraphics[width=50mm]{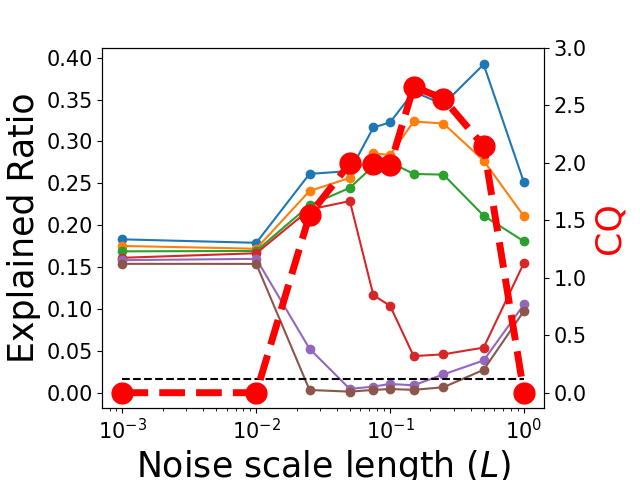} &  \includegraphics[width=50mm]{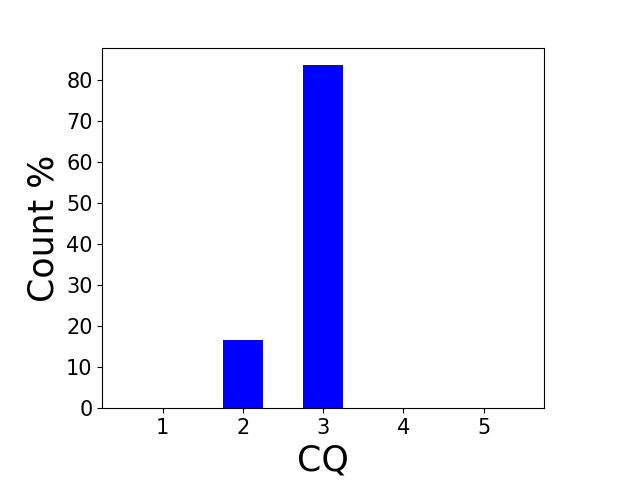} &  \includegraphics[width=50mm]{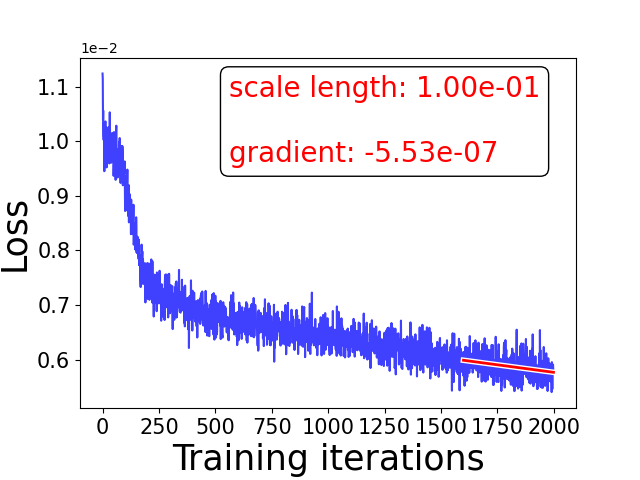} \\
    \end{tabular}
    
\medskip  
Example plots from using the ML method on trajectory data taken from a triaxial model.  The left and middle plots are as described by \cite{Liu2021}.  The right plot shows the change in the loss function as the neural net for a given scale (perturbation) length is being trained.  The scale length in this case is $0.1$.  The gradient is calculated from fitting a straight line to the last 20\% of the loss values from the training iterations (the short red line).
\end{figure}

For the ML method, we use the plots described by \cite{Liu2021} to monitor the behavior of the method (see their explained ratio diagrams), and add a further plot to examine training loss reduction and convergence.  Figure \ref{fig:mlplots} shows example plots taken from our work. 
In PCA terms, explained variance is the variance attributed to individual components, and the ratio is this variance as a fraction of the total variance across all components.  As set out in \cite{Liu2021}, low ratios for a component for a number of consecutive scale factors are taken to indicate a conserved quantity (component), and high ratios unconserved quantities.  The valley \cite{Liu2021} see around a scale factor of $0.1$, we also see in our results.  Their ratio cut-off of $0.1/N$ is not met consistently in our results for 6D data modeling (in particular, in our axisymmetric and triaxial galaxy models) but is met when modeling with 4D data.  As a consequence we treat the explained ratio diagrams as informative but prefer to quote results based on $CQ$ estimation as described in the text surrounding the cosine formula in equation (3) of \cite{Liu2021}.  The key difference between the  two mechanisms is that the explained ratio diagrams are based on perturbing a single data point from a trajectory while the $CQ$ estimation uses multiple data points.  In our case, we use $200$ data points perturbed $2000$ times.

\subsection{Galaxy Models}
\label{sec:toys}

In this section, we cover briefly the theoretical galaxy models and potentials we use in creating the orbit trajectories. We include any parameter values required to use the models, and indicate the number of isolating integrals expected.  In what follows $(x, y, z)$ are Cartesian coordinates and $r$ is the spherical radius.  For the axisymmetric model, $(R, z)$ are taken from cylindrical polar coordinates aligned along the symmetry axis of the galaxy. For all models, the total mass $M = 3$, and the gravitational constant $G = 1$.  All units are theoretical.

For the spherical \cite{Plummer1911} galaxy model, the potential $\phi$ is
\begin{equation}
	\phi(r) = - \frac{GM}{\sqrt{r^2 + b^2}},
\end{equation}
where $b$ is the scale length ($b = 0.5$ in our models). 
For the \cite{Hernquist1990} model, 
\begin{equation}
	\phi(r) = - \frac{GM}{r + a},
\end{equation}
where $a$ is the scale length ($a = 0.5$ in our models).  For both potentials, we expect four integrals to be counted by our modeling, energy and the three components of angular momentum.  Given that angular momentum is conserved, an orbit's trajectory will be in a plane orthogonal to the angular momentum vector.  In such an orbit plane, trajectories have a rosette appearance, for example as in figure 3.1 in \cite{BT2008} and here in our Figure \ref{fig:resplumcq3}, and are confined to the region between an inner and an outer circle.

For the axisymmetric logarithmic models based on \cite{Richstone1980},
\begin{equation}
	\phi(R, z) = \frac{1}{2} v_0^2\log (R^2 + \frac{z^2}{q^2} + R_c^2),
\end{equation}
where $v_0$ is the scale velocity, $q$ is the $z$-axis flattening, and $R_c$ is the core radius. In our models, $v_0 = 1$ and $q = 0.9$, and $R_c = 0.1$ or $R_c = 0$ depending on whether a cored or singular (uncored) model is being used.  We expect three integrals (energy, the z component of angular momentum, and a `non-classical' third integral) to be counted for models in 3D Cartesian coordinates, but only two (energy and the third integral) for 2D meridional plane $(R, z)$ coordinates.  Orbit trajectories are typically tube orbits about the z-axis, for example, as in figure 3.4 in \cite{BT2008}.

For the triaxial, perfect ellipsoid galaxy model \citep{deZeeuw1985}, 
\begin{equation}
	\phi(x, y, z) = - \frac{GM}{\pi} \int_0^{\infty}\frac{1}{1 + m^2(u)}\frac{du}{\sqrt{(a^2 + u)(b^2 + u)(c^2 + u)}},
\end{equation}
where
\begin{equation}
	m^2(u) = \frac{x^2}{a^2 + u} + \frac{y^2}{b^2 + u} + \frac{z^2}{c^2 + u},
\end{equation}
with $a \geq b \geq c > 0$.  For consistency with table 2 of \cite{deZeeuw1985}, we take $a = 1$, $b = 0.625$, and $c = 0.5$.  The relevant equations in \cite{deZeeuw1985} are, for ellipsoidal coordinates, equations 6 to 10, for the potential, and equations 13 to 17 and section 4.2 for the integrals of motion.  We expect three integrals of motion to be counted by our modeling, energy and two related to angular momenta. These integrals are consistent with the St\"{a}ckel formulation of the perfect ellipsoid potential.  Depending on the values of the integrals, orbit trajectories are associated with one of four families - boxes, inner long axis tubes, outer long axis tubes, and short axis tubes as described in \cite{deZeeuw1985} section 5.1.   We adopt the same integration style for the potential and its Cartesian coordinate derivatives as in section 2 of \cite{Merritt1996}, where a substitution is used to make integration bounds finite.  Although trajectories in ellipsoidal coordinates are unlikely observationally, we do use them in some of our models for completeness.

\subsection{Orbit Initial Conditions and Trajectories}
\label{sec:ics}

The same initial conditions (spatial positions and velocities) for the trajectories are used for each galaxy, regardless which modeling method is being utilized.  For the spherical models, we create a spatial distribution of positions that matches the theoretical mass density distribution associated with the galaxy, and use Gaussian sampling with mean zero and the theoretical velocity dispersion to allocate velocity values.  For the axisymmetric model, we use a three isolating integral scheme as employed in, for example,  \cite{Cappellari2006} or \cite{Long2012}.  For the triaxial model, we match its density distribution to create spatial positions and then allocate velocities uniformly randomly having previously assigned kinetic energies to the initial spatial positions of the orbits. In creating the initial conditions, no attempt has been made to create orbits with specific characteristics: we have just reused our existing stellar dynamical modeling software.

Orbit trajectories are created in a given galaxy potential using a 3rd order leapfrog integrator \citep{Ruth1983} with a time step of $10^{-2}$ units.  With such a scheme, the known isolating integrals are reproduced to approximately 1 part in $10^{7}$.   The number of trajectory points varies according to the orbital period of an orbit.  Orbital periods are determined for spherical and axisymmetric models using an epicyclic approximation  \citep[see][section 3.2.3]{BT2008} with an equivalent long axis tube orbit approximation being employed for the triaxial model \citep{Valluri1998}.  Based on our experiences, we limit orbit trajectories to $75$ orbital periods \citep[][used $50$]{Barnes2001}, and set an overall upper limit on the number of trajectory points per orbit to $640K$. Using more orbital periods has little impact on our results.  To avoid any short distance bias to the modeling calculations only every nth point is actually utilized.  From experimentation, we take $n=16$ for triaxial models, $n=8$ for spherical models, and $n=4$ for axisymmetric models.   Trajectories are always constructed using Cartesian coordinates, and then converted later to other coordinate systems as required by the modeling.

The software base for initial conditions and trajectories is taken from the lead author's implementation of the \cite{Syer1996} made-to-measure stellar dynamical modeling method and the \cite{Schwarz1979} orbit based modeling method.  This software was first used in \cite{Long2010}, and most recently in \cite{Long2021}.

\subsection{Resonances}
\label{sec:res}
Resonances are concerned with integer relationships between the main frequencies of the coordinate components of an orbit's trajectory.  A resonance is defined by
\begin{equation}
\label{eqn:res}
	\sum _k ^K m_kw_k = 0,
\end{equation}
where the $w_k$ are the main frequencies, $m_k$ are integer coefficients, and $K$ is the dimension of configuration space ($2$ or $3$ in our work).  We determine component frequencies using the Python NAFF implementation by \cite{Zis2019} (NAFF is the Numerical Analysis of Fundamental Frequencies described in \cite{Laskar1990, Laskar1993}).  We will not describe NAFF in any detail: there is plenty of material regarding its theory and operational use in the literature, for example, \cite{Papa1996, Papa1998}, \cite{Valluri1998}, \cite{Merritt1999}, and \cite{Wang2016}.  We concentrate on describing how we provide input to NAFF, and use its output to find resonances.  There is an alternative approach in \cite{Carpintero1998} for obtaining resonances which we do not utilize.

We provide input to NAFF in complex form ($z_k$) combining component positions ($x_k$) as the real part and velocities ($v_k$) as the imaginary part,
\begin{equation}
	z_{k,j} = x_{k,j} + iv_{k,j},
\end{equation}
where subscript $j$ indicates the jth point along the orbit's phase space trajectory.  The outputs from NAFF relevant to our purposes are the component amplitudes (the real amplitudes, to be precise) and their frequencies.  Practically, it is convenient to sort the component amplitudes and their frequencies into descending amplitude order to facilitate the searching for frequencies that is required.  Quite how we select the main frequencies depends on whether we are working with 2D or 3D orbits. By default, we use 3D orbits but there are times when using 2D alternatives are appropriate, for example, when we are working with orbital plane or meridional plane coordinates.

For 2D orbits, the first component's main frequency ($w_1$) is taken as the frequency associated with the first component's maximum amplitude.  Searching the second component frequencies (searching as described above), the second main frequency ($w_2$) is taken as the first encountered frequency that is not equal to first component's main frequency ($w_1$).

For 3D orbits, the first main frequency ($w_1$) is taken as the maximum amplitude frequency from the component that has the largest amplitude across all three components.  The second main frequency ($w_2$) comes from the component that has the second largest amplitude across all three components, and is the frequency (after searching) that does not have an integer relationship with the first main frequency ($w_1$).  The third main frequency ($w_3$) comes from the remaining component, and is the frequency that does not have an integer linear relationship with the two main frequencies previously determined ($w_1$ and $w_2$).

The final step is to determine whether or not integer coefficients ($m_k$) can be found such that the main frequencies ($w_k$) have an integer relationship as in Equation (\ref{eqn:res}).  We do this using the integer programming capabilities of the publicly available Python package OR-Tools (github.com/google/or-tools).  If we are able to find such integer coefficients, then we have found a resonance.

\section{Results}
\label{sec:results}
We group our results by galaxy model, and include results from both modeling methods in the same figures for ease of comparison.  In these plots and figures, results are color-coded with blue being used for the CI method, and orange for the ML method.  Most of the comparison plots are displaying percentages.  With 64 orbits in total for each galaxy model, $1$ orbit equates to  $\approx 1.6\%$ of the total.  For individual orbits, we show typical results plots in Figures \ref{fig:ciplots} and \ref{fig:mlplots}, in Sections \ref{sec:CImethod} and \ref{sec:MLmethod}.  The galaxy models we have employed mean that we know the number of isolating integrals that should be found.  Ignoring any non-isolating integrals and resonances, this gives us a basis for examining our results, and enables us to discuss whether more or less conserved quantities than expected have been detected by the methods.

\subsection{Spherical Galaxy Models}
\begin{figure}[h]
    \centering
    \caption{Spherical Galaxy Models - Plummer and Hernquist}
	\label{fig:resplumhern}
    \begin{tabular}{cM{65mm}M{65mm}M{65mm}}   
    		& \textit{Plummer} & \textit{Hernquist} \\		
      	\textit{3D} & \includegraphics[width=65mm]{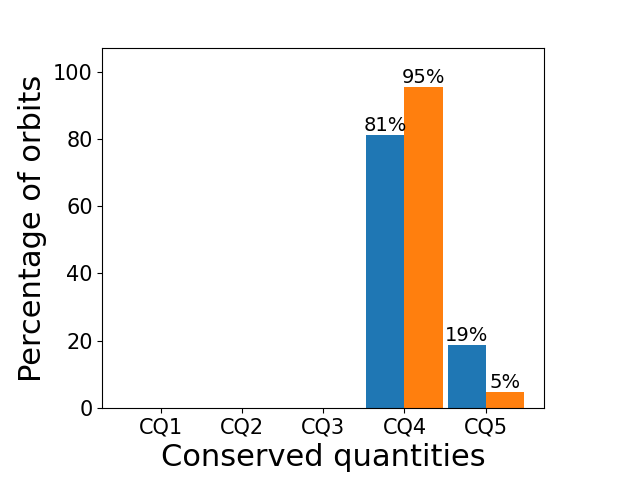} &  \includegraphics[width=65mm]{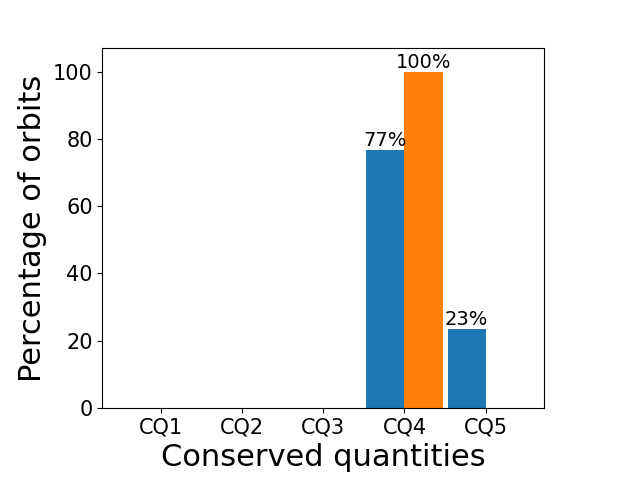}   \\
      	\textit{2D} & \includegraphics[width=65mm]{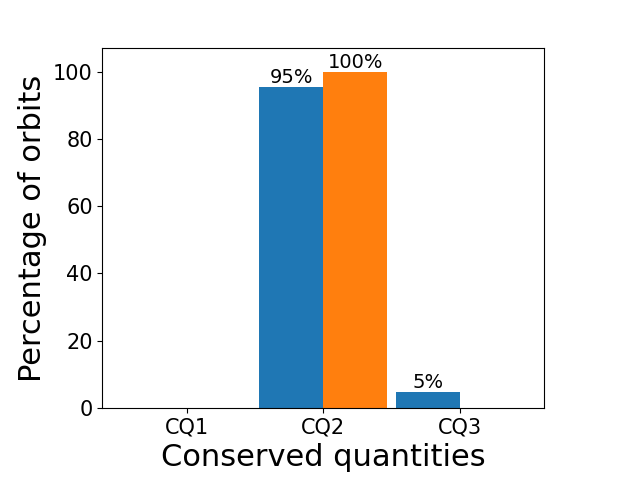} &  \includegraphics[width=65mm]{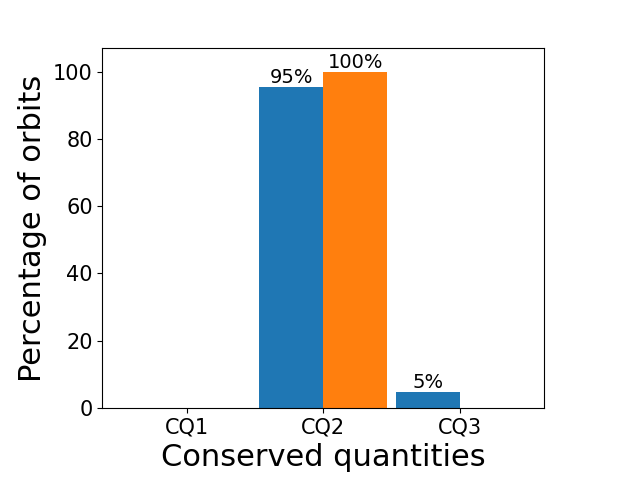}   \\
    \end{tabular}
    
\medskip  
Plummer and Hernquist models in 3D Cartesian coordinates and 2D orbital plane coordinates. The plots show histograms of the percentages of orbits having the indicated numbers of conserved quantities. As stated at the start of Section \ref{sec:results}, CI results are in blue and ML results in orange.  We expect the peak percentages to be associated with $CQ=4$ for 3D modeling and with $CQ = 2$ for 2d modeling, and this is met in practice.
\end{figure}

Results for the Plummer and Hernquist spherical galaxy models, using both the CI and ML methods with 2D and 3D configuration spaces, are shown in Figure \ref{fig:resplumhern}.  

\begin{figure}[h]
    \centering
    \caption{Plummer `Thin' Ring $CQ = 5$ Orbits}
	\label{fig:resplumcq5} 
	\begin{tabular}{ccc}   			
     	\includegraphics[width=50mm]{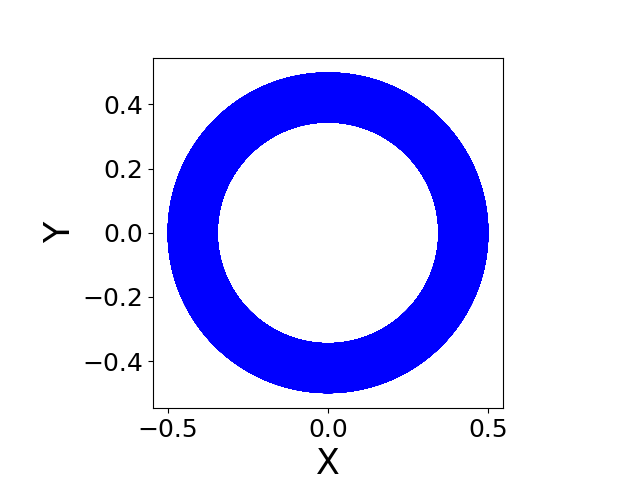} & \includegraphics[width=50mm]{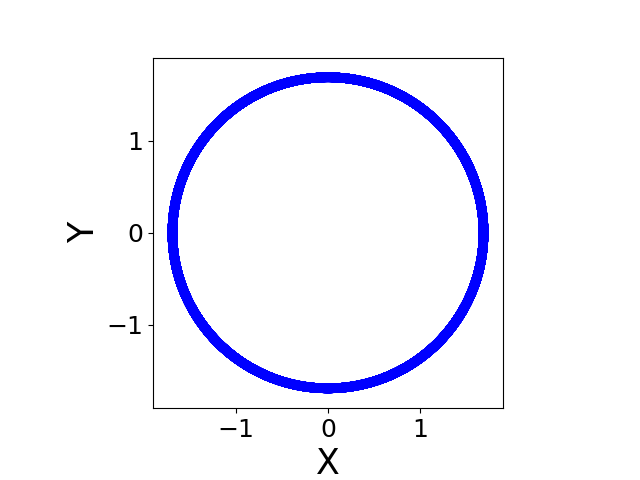} & \includegraphics[width=50mm]{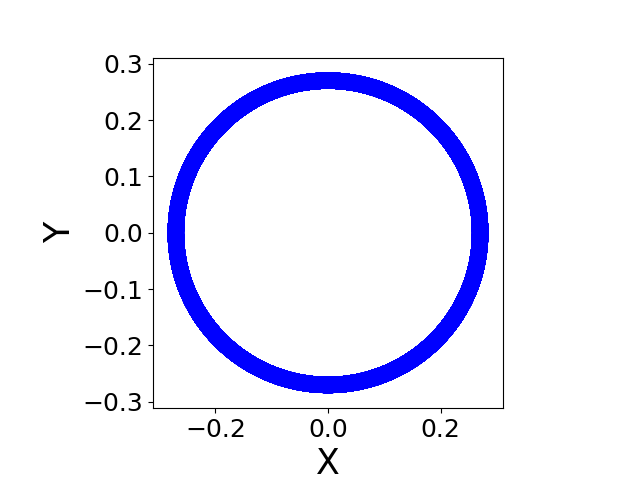}\\
    \end{tabular}
    
\medskip  
Plummer model `thin' ring $CQ = 5$ orbits plotted in their orbital planes.  Both the CI and ML methods agree on the categorization of these orbits as $CQ = 5$.  As such the orbits have more conserved quantities than anticipated.  In this case the categorization appears to be as a result of the orbits having a non-isolating integral of motion.
\end{figure}

Considering first the Plummer model 3D results, they are largely as anticipated with all the orbits having either $CQ = 4$ or $CQ = 5$ conserved quantities.  The CI and ML methods achieve the same CQ for $86\%$  (55 out of 64) of orbits. Where the methods differ is in the number of orbits with more conserved quantities than anticipated ($CQ = 5$), with 12 for the CI method and 3 for the ML method.  The 3 ML orbits are found by the CI method as well.  This difference ($12$ vs $3$ orbits) may be related to whether or not the methods are able to handle non-isolating integrals as conserved quantities. Certainly, `ring' orbits, where the orbital plane inner radius is approximately equal to the outer radius, are present in the orbit sample.  Such orbits are described as having non-isolating integrals in \cite{BT2008} section 3.1, equation 3.62.  The 3 common $CQ = 5$ orbits are `thin' ring orbits (see Fig. \ref{fig:resplumcq5}) whereas the additional 9 orbits found by the CI method have slightly `fatter' rings.

\begin{figure}[h]
    \centering
    \caption{Plummer 2D Orbital Plane - Orbits with additional CQs}
	\label{fig:resplumcq3} 
	\begin{tabular}{ccc}   			
     	\includegraphics[width=50mm]{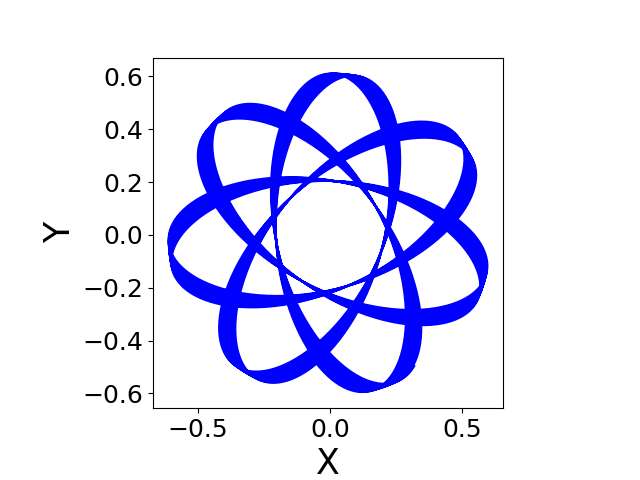} & \includegraphics[width=50mm]{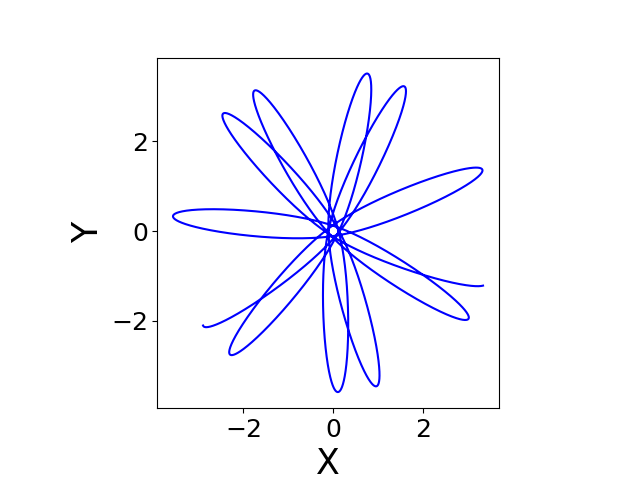} & \includegraphics[width=50mm]{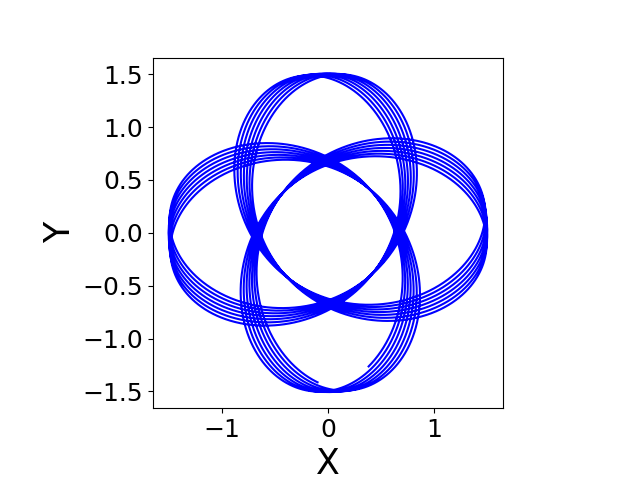}\\
    \end{tabular}
    
\medskip  
Plummer model 2D orbital plane orbits having more conserved quantities ($CQ > 2$) than expected.  The orbit trajectories have been restricted to 10K data points so that the orbit structures are visible.
As may be seen, none of the orbits has a `ring' structure. Without formulae for the conserved quantities, it is not clear why the CI method has determined the orbits have $CQ = 3$ rather than the anticipated $CQ = 2$.
\end{figure}

Almost all the 2D orbital plane Plummer orbits have the expected number of conserved quantities ($CQ = 2$) with only 3 orbits being identified by the CI method as having more quantities ($CQ = 3$).  None of these orbits is part of the $CQ = 5$ set of orbits from the 3D work above.  
Also, as can be seen from Figure \ref{fig:resplumcq3}, none of the orbits has an obvious `thin' ring structure.  Without having a mechanism to attempt to determine the formulae for the conserved quantities, it is not clear why the CI method identified these orbits as having more conserved quantities. 
Similarly, it is unclear why the ML method has not identified any such orbits as it was able to do when modeling in 3D.

The Hernquist model results, both 2D and 3D, are very similar to the Plummer results.  However, the ML method for the Hernquist model finds no orbits with more conserved quantities than anticipated, that is, no orbits with $CQ = 5$ (3D) nor with $CQ = 3$ (2D).  The CI method is able to find $8$ $CQ = 5$ orbits with a ring structure, of which $5$ are considered to be `thin'.  In other words, the CI method is able to identify at least some orbits as having non-isolating integrals.  The ML method appears to find none for the Hernquist models but was able to do so for the Plummer model.

\subsection{Axisymmetric Galaxy Models}
\begin{figure}[h]
    \centering
    \caption{Axisymmetric Logarithmic Models - Singular and Cored}
	\label{fig:resaxi}
    \begin{tabular}{cM{65mm}M{65mm}M{65mm}} 
    			& \textit{Singular} & \textit{Cored} \\
    		\textit{3D}  & \includegraphics[width=65mm]{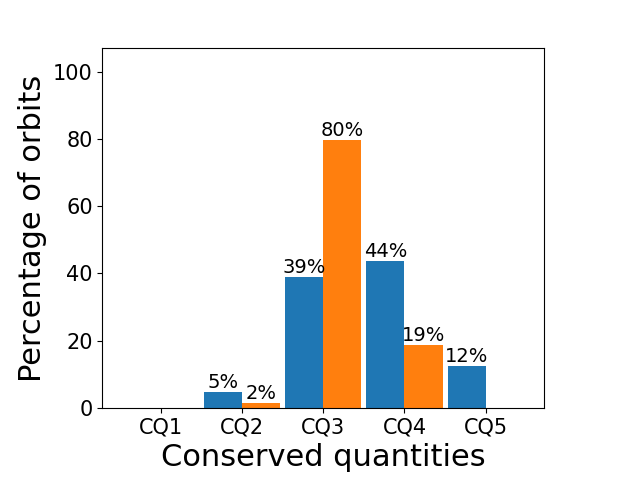} & \includegraphics[width=65mm]{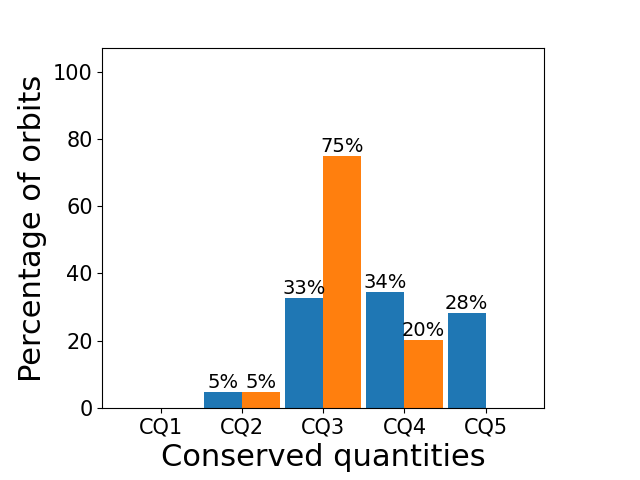} \\
    		\textit{2D} & \includegraphics[width=65mm]{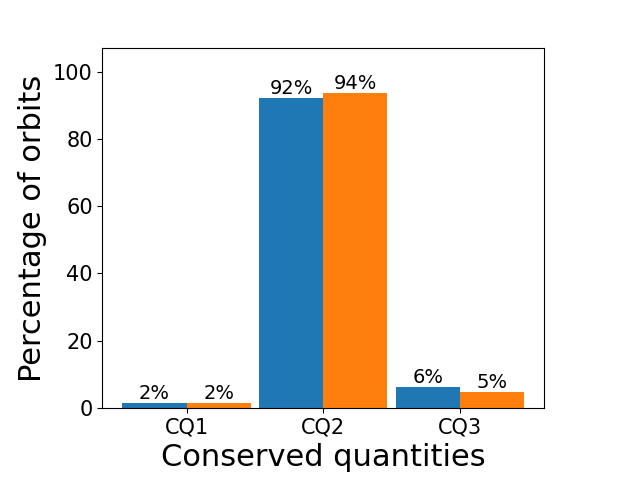} & \includegraphics[width=65mm]{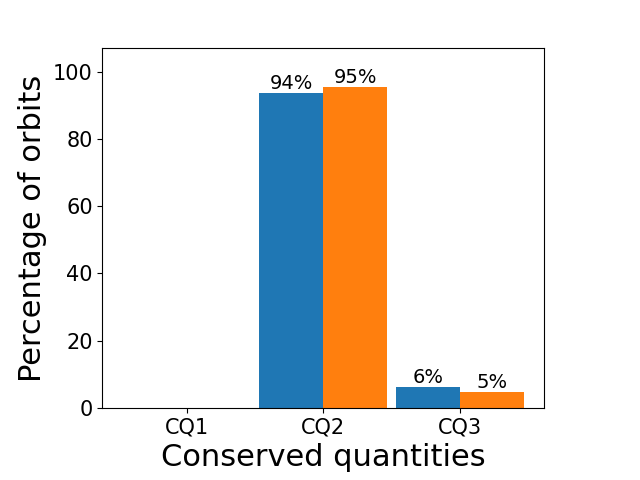} \\  	
    \end{tabular}
    
\medskip  
Singular and cored axisymmetric logarithmic models in 3D Cartesian coordinates and 2D meridional plane coordinates.  We expect the peak percentages in the histograms to be associated with $CQ=3$ for 3D modeling and with $CQ = 2$ for 2d modeling.  For the 3D CI models, this peak expectation is not met.
\end{figure}

Results for the singular and cored axisymmetric logarithmic galaxy models, using both the CI and ML methods with 2D and 3D configuration spaces, are shown in Figure \ref{fig:resaxi} and Tables \ref{tab:axiresuncore} and \ref{tab:axirescore}.

\begin{table}[h]
	\centering
	\caption{Axisymmetric Logarithmic Models - Singular}
	\label{tab:axiresuncore}
		\begin{tabular}{cc|cc|cc|cc}
		\hline
		 & \textbf{CQ} & \multicolumn{2}{c|}{\textbf{CI Method}} & \multicolumn{2}{c|}{\textbf{ML Method}} & \multicolumn{2}{c}{\textbf{Common}} \\
		 &    & Orbits & Resonances & Orbits & Resonances & Orbits & Resonances \\
		\hline 
			& 1 & 0  & 0  & 0  & 0  & 0  & 0\\
			& 2 & 3  & 2  & 1  & 1  & 1  & 1\\
		\textit{3D}  & 3 & 25 & 4  & 51 & 18 & 21 & 3\\
			& 4 & 28 & 11 & 12 & 3  & 8  & 2\\
			& 5 & 8  & 5  & 0  & 0  & 0  & 0\\
		\hline
			& 1 & 1  & 1  & 1  & 1  & 1  & 1 \\
		\textit{2D}  & 2 & 59 & 19 & 60 & 19 & 58 & 19\\
			& 3 & 4  & 2  & 3  & 2  & 2  & 2 \\
		\hline
	\end{tabular}

\medskip
Axisymmetric singular logarithmic models showing the distribution of orbits and resonances by model-determined CQ value.  The \textit{Orbits} and \textit{Resonances} columns give the number counts of orbits, and orbits with resonances, for each $CQ$ value.  The \textit{Common} column gives the number of times the CI and ML methods agree on the $CQ$ categorization.  The main points to note are that some orbits have more conserved quantities than might be expected and this can not be explained by resonances alone; resonances, with the exception of the CI-3D results, do not appear to be identified very well; and, lastly, the 2D meridional results show less variation (lack of discrimination ?) than the 3D results.
\end{table}

The pattern of results is similar between the singular and cored models.  We examine the singular results first and then deal with any differences shown by the cored results.  Looking at the plot of the singular modeling with 3D Cartesian coordinates in Figure \ref{fig:resaxi}, it is clear that orbits have predominately $CQ \geq 3$ (as expected) but that the ML results are more peaked at $CQ = 3$ than the CI results: the CI results have significantly more orbits with additional conserved quantities ($CQ > 3$).  The two methods only achieve the same results for $47\%$ of the orbits (30 out of 64).  Looking at the 2D meridional plane results in Figure \ref{fig:resaxi}, the agreement is much higher with $95\%$ of the orbits (61 out of 64) having the same results but with few orbits having additional conserved quantities.

\begin{figure}[h]
    \centering
    \caption{Singular Axisymmetric Model - Orbits with additional CQs}
	\label{fig:resaxicq3} 
	\begin{tabular}{cc}   			
     	\includegraphics[width=50mm]{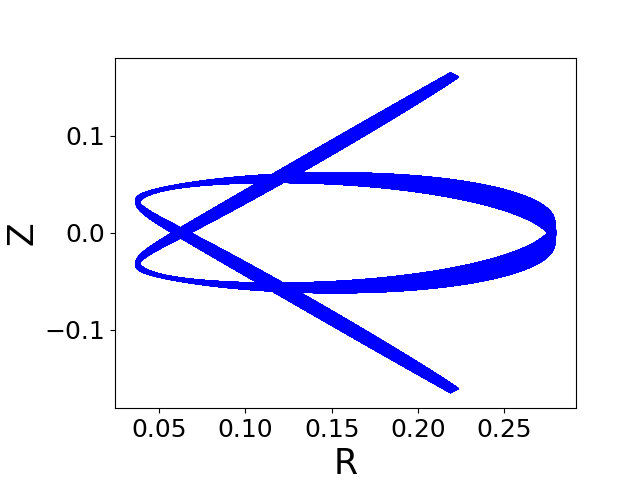} & \includegraphics[width=50mm]{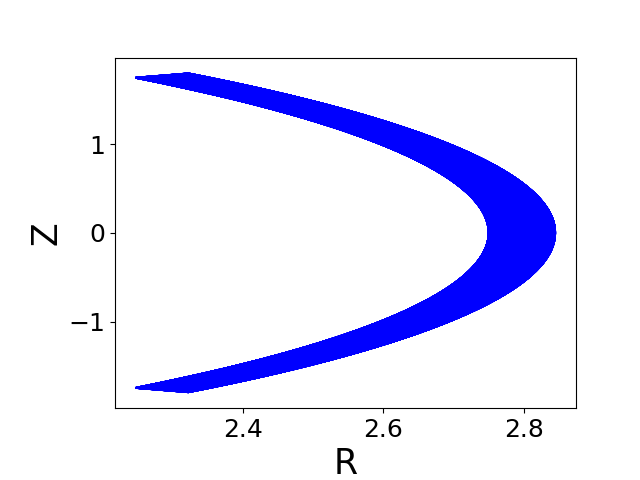} \\
    \end{tabular}
    
\medskip  
Singular axisymmetric model orbits with additional conserved quantities.  Both the orbits are categorized as $CQ = 3$ (2D modeling) and $CQ = 4$ (3D modeling) by both methods.  As such, these orbits have more conserved quantities than might be expected from considering only isolating integrals.  The resonance analysis of the orbits indicates both are resonant.
\end{figure}

A resonance analysis (Sect. \ref{sec:res}) of the singular model orbits shows that $22$ orbits are resonant, and might be expected to have an increased number of conserved quantities ($CQ > 3$).  The CI method 3D Cartesian results have $11$ orbits with $CQ = 4$, and $5$ with $CQ = 5$.  By comparison, the ML method results have only $3$ orbits with $CQ = 4$; of the remaining $19$, $18$ have just the anticipated number of conserved quantities ($CQ = 3$) which seems not be correct.  Examining the 2D meridional plane results, the CI method has $2$ resonant orbits (see Fig. \ref{fig:resaxicq3}) having an increased number of conserved quantities ($CQ = 3$), with the ML method having the same orbits also with $CQ = 3$.   In the 3D results, both methods show  orbits with increased conserved quantities ($CQ = 4$). 

From the previous paragraphs in this section, hopefully it is clear that the numbers of orbits with increased conserved quantities ($CQ \geq 4$) in the 3D singular runs can not be explained by resonances alone.  The numbers imply that an appreciable number of orbits must have non-isolating integrals. Having formulae for the conserved quantities would enable these orbits to be investigated.

\begin{figure}[h]
    \centering
    \caption{Singular Axisymmetric Model - Orbits with less CQs}
	\label{fig:resaxicq2} 
	\begin{tabular}{ccc}   			
     	\includegraphics[width=50mm]{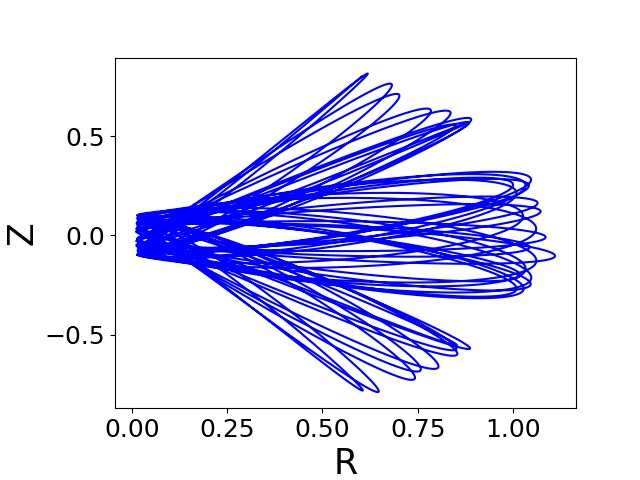} & \includegraphics[width=50mm]{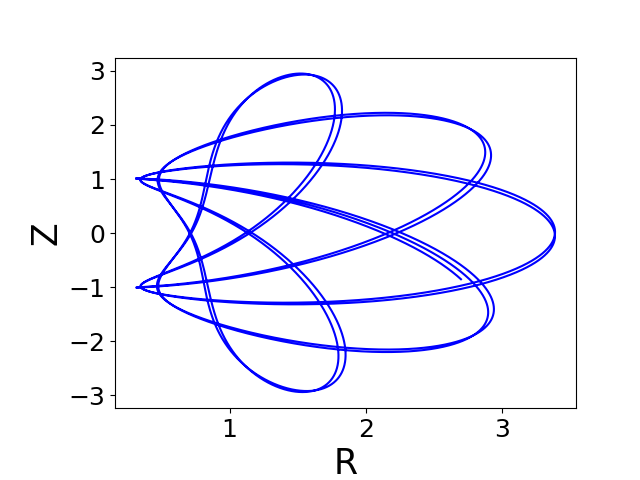} & \includegraphics[width=50mm]{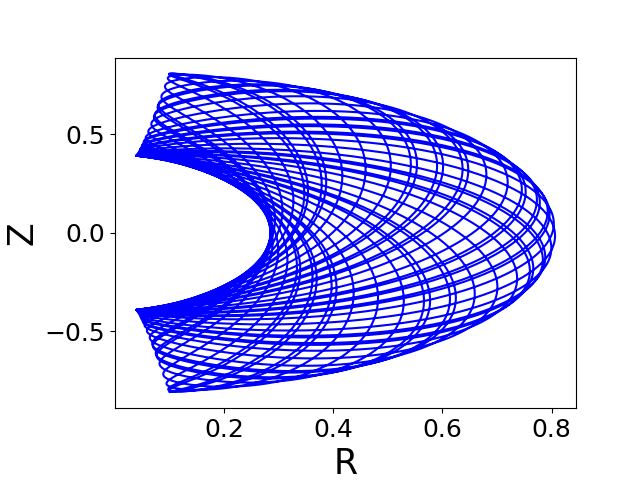}\\
    \end{tabular}
    
\medskip  
Singular axisymmetric model orbits with less conserved quantities. From the 3D modeling, both the CI and ML methods agree these three orbits have less conserved quantities than might be expected from considering only isolating integrals  ($CQ < 3$).   The orbit in the left hand plot has similarities with the chaotic orbit in figure 7 of \cite{Zotos2014}.  The remaining orbits do not appear to be chaotic but may have a reduced number of isolating integrals.  The 2D meridional plane modeling also indicates that the left hand plot orbit has less conserved quantities ($CQ = 1$).
\end{figure}

The $3$ singular 3D model orbits with less conserved quantities ($CQ = 2$) than expected are shown in Figure \ref{fig:resaxicq2}.  The left hand plot orbit trajectory visually appears to be similar to the chaotic orbit in figure 7 of \cite{Zotos2014}, and has both the CI and ML methods agreeing on its categorization.  Also, the orbit is categorized similarly in the 2D meridional plane modeling.  Quite why the other $2$ orbits have category $CQ = 2$ is unclear.

\begin{table}[h]
	\centering
	\caption{Axisymmetric Logarithmic Models - Cored}
	\label{tab:axirescore}
		\begin{tabular}{cc|cc|cc|cc}
		\hline
		 & \textbf{CQ} & \multicolumn{2}{c|}{\textbf{CI}} & \multicolumn{2}{c|}{\textbf{ML}} & \multicolumn{2}{c}{\textbf{Common}} \\
		 &    & Orbits & Resonances & Orbits & Resonances & Orbits & Resonances \\
		\hline 
			& 1 & 0  & 0  & 0  & 0  & 0  & 0\\
			& 2 & 3  & 2  & 3  & 1  & 0  & 0\\
		\textit{3D}  & 3 & 21 & 7  & 48 & 18 & 20 & 7\\
			& 4 & 22 & 9  & 13 & 7  & 8  & 4\\
			& 5 & 18 & 8  & 0  & 0  & 0  & 0\\
		\hline
			& 1 & 0  & 0  & 0  & 0  & 0  & 0 \\
		\textit{2D}  & 2 & 60 & 23 & 61 & 24 & 60 & 23\\
			& 3 & 4  & 3  & 3  & 2  & 3  & 2 \\
		\hline
	\end{tabular}

\medskip
Axisymmetric cored logarithmic models showing the distribution of orbits and resonances by model-determined CQ value. The columns are as defined for Table \ref{tab:axiresuncore}.  Not that unsurprisingly, given the similarities in gravitational potentials, the results are consistent with the singular modeling results in Table \ref{tab:axiresuncore}.
\end{table}

The results for the cored axisymmetric model are very similar to the singular results, so we will focus on just the resonance related matters in Table \ref{tab:axirescore}.  For the 3D cored model, a resonance analysis shows that there are $26$ resonant orbits. $17$ of these orbits ($65\%$) have been categorized by the CI method as having an increased number of conserved quantities ($CQ \geq 4$).  The ML method is only able to categorize $7$ resonant orbits as $CQ = 4$.  More broadly, the CI method has $40$ orbits
with $CQ \geq 4$, while the ML method has $13$ orbits.  Even with the resonant orbits subtracted, there is still an appreciable number left with an increased number of conserved quantities that require explaining.  Having formulae for the conserved quantities would assist matters.

\subsection{Triaxial Galaxy Model}
\begin{figure}[h]
    \centering
    \caption{Triaxial Model - Perfect Ellipsoid}
	\label{fig:restri}
    \begin{tabular}{cc}   		
      	\textit{3D Cartesian Coordinates} & \textit{3D Ellipsoidal Coordinates} \\
      	\includegraphics[width=70mm]{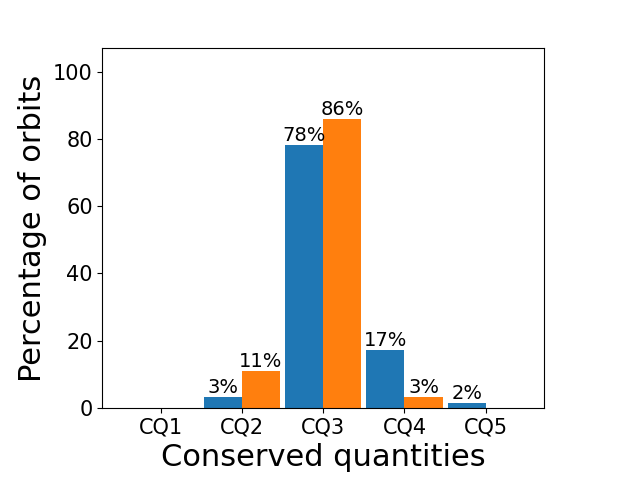}    &  \includegraphics[width=70mm]{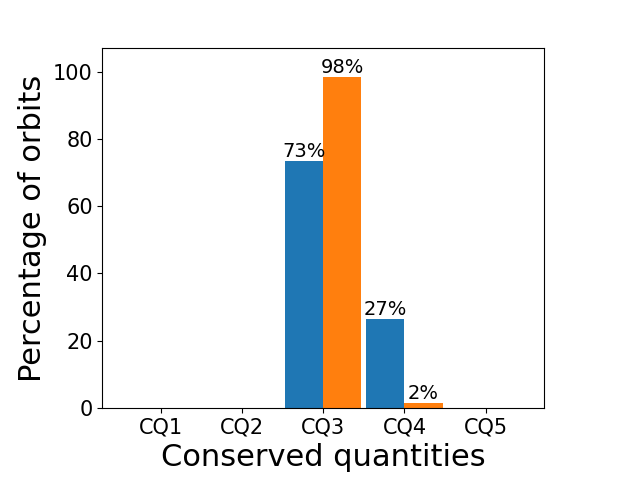}  \\
    \end{tabular}
    
\medskip  
Perfect ellipsoid triaxial models in 3D Cartesian coordinates and 3D ellipsoidal coordinates.  We expect the peak percentages in the histograms to be associated with $CQ=3$, and this is met in practice.
\end{figure}

Results for the perfect ellipsoid triaxial galaxy model, using both Cartesian and ellipsoidal coordinates, are shown in Figure \ref{fig:restri} and Table \ref{tab:trires}.  The 64 orbit sample for this triaxial galaxy comprises 37 X-tube orbits (58\%), 10 Z-tube (16\%), and 27 box orbits (26\%).

\begin{table}[h]
	\centering
	\caption{Triaxial Model - Perfect Ellipsoid}
	\label{tab:trires}
		\begin{tabular}{cc|cc|cc|cc}
		\hline
		 & \textbf{CQ} & \multicolumn{2}{c|}{\textbf{CI}} & \multicolumn{2}{c|}{\textbf{ML}} & \multicolumn{2}{c}{\textbf{Common}} \\
		 &    & Orbits & Resonances & Orbits & Resonances & Orbits & Resonances \\
		\hline 
			& 1 & 0  & 0  & 0  & 0  & 0  & 0\\
			& 2 & 2  & 0  & 7  & 3  & 0  & 0\\
	\textit{Cartesian}    & 3 & 50 & 11 & 55 & 13 & 44 & 9\\
			& 4 & 11 & 6  & 2  & 2  & 2  & 2\\
			& 5 & 1  & 1  & 0  & 0  & 0  & 0\\
		\hline
			& 1 & 0  & 0  & 0  & 0  & 0  & 0 \\
		    & 2 & 0  & 0  & 0  & 0  & 0  & 0 \\
	\textit{Ellipsoidal}	& 3 & 47 & 13 & 63 & 17 & 47 & 13 \\
			& 4 & 17 & 5  & 1  & 1  & 1  & 1\\
			& 5 & 0  & 0  & 0  & 0  & 0  & 0\\
		\hline
	\end{tabular}

\medskip
Perfect ellipsoid triaxial models showing the distribution of orbits and resonances by model-determined CQ value.  The columns are as defined for Table \ref{tab:axiresuncore}. The distribution patterns are consistent with those for the axisymmetric models - see Tables \ref{tab:axiresuncore} and \ref{tab:axirescore}.
\end{table}

The results for the triaxial modeling are generally as anticipated with most orbits having $CQ = 3$ as expected, or $CQ = 4$, an increased number of conserved quantities.  Both methods have categorized some orbits as having fewer conserved quantities than expected ($CQ = 2$) indicating irregular orbits.  

A resonance analysis (Sect. \ref{sec:res}) of the orbits shows that $18$ are resonant. Using 3D Cartesian coordinates, the CI method categorizes $7$  of these orbits in the $CQ = 4$ or $CQ = 5$ bands where they might be expected to be.  By comparison, the ML method only places $2$ orbits in the $CQ = 4$ band.
This would appear to leave $11$ orbits (CI method) and $16$ orbits (ML method) unrecognized as resonant orbits.  The equivalent orbit numbers, unrecognized when ellipsoidal coordinates are used, are $13$ orbits (CI method) and $17$ orbits (ML method).  

The CI and Ml methods reach the same CQ categorization for $72\%$ (46 out of 64) of the orbits when Cartesian coordinates are in use, and $75\%$ (48 out of 64) when using ellipsoidal coordinates.  The $CQ = 2$ orbits, indicating orbits with fewer conserved quantities than expected, are a concern, particularly for the ML method, as all the orbits for the triaxial galaxy model have $3$ isolating integrals by construction (which were explicitly confirmed as conserved during trajectory construction - see Sect. \ref{sec:ics}).
Note that no $CQ = 2$ orbits have been counted when modeling with ellipsoidal coordinates.  Also, it should be noted that the $CQ = 4$ orbits which are not resonant may have non-isolating integrals.
Having formulae for the conserved quantities would enable these unexplained $CQ = 2$ and $CQ = 4$ orbits to be investigated.

\subsection{Computer Utilization}
\label{sec:computil}

Modeling runs were performed on a $20$ core desktop PC.  No attempt was made to use graphics processing units (GPUs) to improve performance but their use is not precluded.  All software used is Python 3 based with some use of Cython for performance critical code. PyTorch is used in the machine learning scripts.  Multi-processor working, limited to a maximum of 10 processors, was invoked for both the CI and ML implementations. 

\begin{table}[h]
	\centering
	\caption{Modeling Runtimes per Orbit}
	\label{tab:perfpo}
		\begin{tabular}{r|cc|cc}
		\hline
		 & \multicolumn{2}{c|}{\textbf{3D Cartesian}} & \multicolumn{2}{c}{\textbf{Alternative}} \\
		 & \multicolumn{2}{c|}{\textbf{Coordinates}} & \multicolumn{2}{c}{\textbf{Coordinates}} \\
		 & CI & ML & CI & ML \\
		\hline 
		\textbf{Spherical Models} &&&&\\
		Plummer   & 1s & 5s & 1s & 5s \\
		Hernquist & 1s & 5s & 1s & 5s \\
		\textbf{Axisymmetric Models} &&&& \\
		Singular &  5s & 38s &  4s & 18s  \\
		Cored   & 10s & 39s & 10s & 18s \\
		\textbf{Triaxial Model} &&&& \\
		Ellipsoid & 8s & 58s & 8s & 28s \\
		\hline
	\end{tabular}

\medskip
Processing runtimes per orbit for the different galaxy models. The alternative coordinates are 2D orbital or meridional plane coordinates for the spherical and axisymmetric models respectively, while for triaxial the coordinates are 3D ellipsoidal coordinates.  Times for the ML method are appreciably higher than the CI method: this is explained in Section \ref{sec:computil} by the need to train neural networks.  
\end{table}

In Table \ref{tab:perfpo}, we show the processing runtimes for the various galaxy models using both methods up to an equivalent point in each method.  For the CI method this point is just after the correlation integral has been computed, and for the ML method just after neural net training has been completed.  
Internally, the ML method trains one neural network per scale length (10 in total for our purposes).  For the CI method, the correlation integral is computed from a single pass of the trajectory data using multiple scale lengths: it does not need to be repeated for each scale length.  Given that both methods use the same number of computer cores ($10$), training costs explain why the ML method takes longer.  Note that the performance achieved is dependent on how the methods are configured: changing various hyperparameters (for example, the number of hidden neural net layers, or the number of orbital periods) will change the performance profiles.
In looking at the table, it should be understood that we have used only a few orbits (64) per galaxy model.  If we were looking to analyze galaxy orbits from, say, a cosmological simulation, we might want to analyze several thousand in a reasonable time frame of perhaps just a few hours (less than 8 hours, for example).

\section{Discussion}
\label{sec:discussion}

As will be clear from preceding sections,  our research would benefit from being able to use robust methods and software to determine algebraic formulae for all conserved quantities in a computationally cost effective and timely manner. The key here is \textit{all}: software which will nominally find a \textit{single} conserved quantity does exist (but does not always perform in a timely manner). A very recent article \cite{Liu2022} appeared once the work here was substantially complete and may suggest a solution. Given the problem domain in which we are operating (galaxies), we are able to provide support to any symbolic regression packages trying to find algebraic formulae from trajectory data.   For example, once a coordinate system has been determined for a galaxy, potentially conserved quantities such as momenta can readily be calculated from trajectory coordinates; total energy cannot be, but the kinetic component can.  From the literature, material from the 1990s on genetic algorithms, capable of identifying multiple solutions, is readily identifiable.  A further consequence of the lack of a symbolic regression capability is that no false positive or negative analyses have been possible.

Claims in \cite{Liu2021} that the ML method outperforms the correlation integral (fractal) method do not appear to have been confirmed by our work.  In our particular, galaxy context, the ML method is slower than the CI method.

Resonances are not handled well by either method though the CI method does appear to outperform the ML method, unless of course there is some coincidentally hidden factor in operation.  For non-isolating integrals, they do seem to be recognized for spherical models but the position is much less clear for our axisymmetric and triaxial models.  It may be that spherical model non-isolating integrals are easy to detect because of their simple ring structure but that more complex structures are inherently harder to detect.
What is clear is that for orbits with more than the expected number of conserved quantities (based on counting isolating integrals), the increase can not be explained by resonances alone: it does appear as if there is some other factor contributing which could be non-isolating integrals.
Also, it must not be forgotten that functions of conserved quantities are also conserved.

We have previously commented on increasing the number of hidden neural network layers and their nodes as complexity rises from spherical to triaxial orbit modeling.  Learning rate is another parameter to experiment with but we found no benefit from so doing.  Our approach to setting hyper-parameters has been to assume that using the same parameter values for all orbits from a given galaxy model is acceptable, but it may not be.  It may be appropriate to set parameters for individual orbits, for example, for orbits that remain very close to the center of the galaxy.

It has been useful to use 2D orbits in this research to help examine the behaviors of the methods.  When working with observed Galaxy data or with cosmological simulations, say, it is likely that 3D orbits will be more usual. Related to the orbits is the number of orbit trajectory points available. What would be appropriate in the future would be to investigate how few trajectory points can be used both to count conserved quantities and to determine their formulae. 

Not withstanding the issues we have identified, it may be worth considering using the CI and ML methods in this paper as part of a set of phase space trajectory tools.  For example, such a tool set might comprise
\begin{itemize}
\item the CI and ML methods to count CQs;
\item a trajectory-based angular momentum conservation analyzer;
\item resonance analysis as per Section \ref{sec:res};
\item kinetic energy is not conserved but is easy to calculate from the trajectory for use elsewhere; and
\item using symbolic regression to find the total energy, subtracting the kinetic energy to give the gravitational potential.
\end{itemize}
Work on the last point above is already well advanced and will be submitted for publication in the near future.

\section{Conclusions}
\label{sec:conclusions}
We have met the objectives set out in the Introduction, Section \ref{sec:intro}, in that we have compared two schemes for determining the number of integrals and resonances.  One scheme utilizes machine learning while the other using more traditional methods does not.  

Overall, the results are mixed - neither scheme appears to be clearly better than the other in all aspects.  Both fail to deal with resonances adequately though the correlation integral approach does appear to perform somewhat better.  It is tempting to say that the machine learning approach is better able to identify isolating integrals but this would have to be in the absence of any non-isolating integrals and resonances.  Equally, it could be said that the correlation integral approach appears to be finding some non-isolating integrals but it is difficult to know truly whether they are non-isolating integrals and whether all such integrals have been found.

The reason we say 'appears to' above is that no robust mechanism yet exists for determining the algebraic formulae for multiple conserved quantities  from a single trajectory in a timely manner, though it may be that the investigations of \cite{Liu2022} will suggest a viable solution. Addressing this symbolic regression issue needs to be tackled with some urgency, and this or a similar investigation repeated.

While the issues we have noted are being addressed, the CI and ML methods might form part of a trajectory tool kit for simulated galaxies.

\begin{acknowledgements}
We thank the referee for their comments which have helped improve the paper.  This work is partly supported by the National Key Basic Research and Development Program of China (No. 2018YFA0404501 to Shude Mao), and by the National Science Foundation of China (Grant No. 11821303, 11761131004 and 11761141012 to Shude Mao).  Yougang Wang acknowledges support by the CAS Interdisciplinary Innovation Team (JCTD- 2019-05).
Thanks are due to Ziming Liu for making the AI Poincare software publicly available.
\end{acknowledgements}

\bibliographystyle{raa}
\bibliography{ms2022-0407}

\end{document}